\documentclass[a4paper,conference]{IEEEtran}
\usepackage{cite}
\usepackage{graphicx}
\usepackage{amsmath}
\usepackage{amssymb}
\usepackage{algorithm}

\begin{document}

\title{Compressed Sensing with Incremental Sparse Measurements}

\author{Xiaofu Wu\IEEEauthorrefmark{1}, \IEEEauthorblockN{Zhen Yang\IEEEauthorrefmark{1} and
Lu Gan\IEEEauthorrefmark{2}} \\
\IEEEauthorblockA{\IEEEauthorrefmark{1}
Nanjing University of Posts and Telecommunications, Nanjing 210003, CHINA\\ Email: xfuwu@ieee.org, and yangz@njupt.edu.cn}
\\ \IEEEauthorblockA{\IEEEauthorrefmark{2}Brunel University, London UB8 3PH, UK\\ Email: lu.gan@brunel.ac.uk}}



\maketitle

\begin{abstract}
This paper proposes a verification-based decoding approach for reconstruction of a sparse signal with incremental sparse measurements.
In its first step, the verification-based decoding algorithm is employed to reconstruct the signal with a fixed number of sparse measurements.
Often, it may fail as the number of sparse measurements may be not enough, possibly due to an underestimate of the signal sparsity. However, we observe that even if this first recovery fails, many component samples of the sparse signal have been identified. Hence, it is natural to further employ incremental measurements tuned to the unidentified samples with known locations. This approach has been proven very efficiently by extensive simulations.
\end{abstract}


\IEEEpeerreviewmaketitle

\section{Introduction}
\PARstart{C}{ompressed} Sensing (CS) has received much attention in recent years
\cite{DonohoCS,GilbertCS,XuExpanderCS,ZhangCC_CS,ZhangCSIT}. For a sparse signal, it essentially
does not require to sample the signal with the traditional Nyquist rates.
Basically, one can undersample a sparse signal with a much lower rate determined by the sparsity of the signal and the idea of which stems
from various linear-transform based compression approach \cite{DonohoCS}.

Formally, the CS problem in the noiseless setting considers the estimation
of an unknown and sparse signal vector $\mathbf{e} \in \mathbb{R}^N$ from
a vector of linear observations $\mathbf{s} \in \mathbb{R}^M$, i.e.,
\begin{equation}
  \label{eq:sys}
   \mathbf{s} = H \cdot \mathbf{e},
\end{equation}
where $H \in \mathbb{R}^{M\times N}$ is a fixed matrix
known as measurement matrix and only a small number
(the sparsity index), $K<<N$, of elements of $\mathbf{e}$ are non-zero.
The set containing the positions of these elements is known as the support set, defined as $\mathcal{S}=\left\{i\in [1,N]:  e_i \neq 0\right\}$,
with cardinality $|\mathcal{S}|=K$.

The solution to this system of equations is known to be given by the vector that minimizes $\|\mathbf{e}_0\|_0$
 ($l_0$-norm) subject to $\mathbf{s} = H \cdot \mathbf{e}_0$, which is a non-convex
optimization problem. In \cite{DonohoCS}, it was  established that the vector $\mathbf{e}_1$ with
minimum $l_1$-norm subject to $\mathbf{s} = H \cdot \mathbf{e}_1$ coincides with $\mathbf{e}_0$
whenever the measurement matrix satisfies the well-known restricted isometry property (RIP) condition.
However, it was indicated in \cite{YihongCS} that $l_1$-based reconstruction algorithms are non-optimal.

Recently, there are increased interests in employing sparse measurement matrices and message-passing reconstruction algorithms for CS.
With bipartite-graph representations of sparse measurements, various message-passing algorithms originally developed for decoding sparse-graph codes
have been introduced for reconstruction of a sparse signal\cite{SarvCS,XuExpanderCS,ZhangCSIT,BinGraphCS}. It was shown that the message-passing
decoding algorithms can outperform $l_1$-based theoretical limits. The connections between Low-Density Parity-Check (LDPC) codes and CS
are addressed in detail in \cite{LdpcCS}.

In this paper, we address the problem of the verification-decoding-based reconstruction of a sparse signal from sparse measurements. We observed that whenever the verification decoder fails, the number of unidentified component samples of the sparse signal, however, may be greatly reduced compared to the signal length. Therefore, it is possible to develop an efficient incremental recovery approach based on this observation.
\section{LDPC Coding and Compressed Sensing}
\subsection {LDPC Codes and Syndrome Decoders}
Let $C$ be an $(N, K)$ LDPC code of block length $N$ and dimension
$K$, which has a parity-check matrix $H=[h_{m,n}]$  of $M$ rows, and
$N$ columns. For LDPC codes over $GF(q)$, the Galois Filed of size $q$, it means that $h_{m,n}\in GF(q)$.
If the size of field $q$ is larger than 2, we call them non-binary LDPC codes.

The Tanner graph of an LDPC code is with one-to-one correspondence
with the parity-check matrix.  Let $G=(\mathcal{V}\bigcup \mathcal{C}, \mathcal{E})$ be the
Tanner graph of a code $C$ with respect to the
parity-check matrix $H=[h_{m,n}]$, where the set of variable nodes
$\mathcal{V}$ represents the codeword bits (or columns of $H$) and the set of
check nodes $\mathcal{C}$ represents the set of parity-check constraints (or
rows of $H$) satisfied by the codeword bits.

Throughout this paper, we denote the set of variable nodes that participate in check $m$ by
$\mathcal{N}(m)=\{n: h_{m,n}\neq 0\}$. Similarly, we denote the set of
checks in which variable node $n$ participates as $\mathcal{M}(n)=\{m:
h_{m,n}\neq 0\}$. For each $n \in \mathcal{V}$ and $m \in \mathcal{C}$, let $d_v(n)$ and $d_c(m)$ denote
degrees of variable node $n$ and check node $m$, respectively.

Now consider that a codeword is transmitted over a $q$-ary symmetrical channel.
A codeword $\mathbf{c} \in F^{N}$ in $C$ (over the field $F=GF(q)$) should admit the parity-check constraint of $H \cdot \mathbf{c}=\mathbf{0}$.
At the receiver, the received vector  $\mathbf{\hat{c}}$ can be written as  $\mathbf{\hat{c}} = \mathbf{c} + \mathbf{e}$, where
$\mathbf{e} \in F^{N}$ is the additive noise. With the received vector $\mathbf{\hat{c}}$, the decoder tries to estimate
$\mathbf{c}$ (or equivalently $\mathbf{e}$). A syndrome decoder works by first calculating the syndrome
$\mathbf{s} = H \mathbf{\hat{c}} = H(\mathbf{c} + \mathbf{e}) = H \mathbf{e}$ and then finding the most
likely error patten $\mathbf{\hat{e}}$ given the syndrome $\mathbf{s}$. For the $q$-ary symmetric channel, the syndrome decoder
is a minimum Hamming distance decoder.

For general linear block codes over $GF(q)$, the optimal syndrome decoder often has heavy complexity especially for large fields.
For LDPC codes, there are various low-complexity but suboptimal syndrome decoders, which make a clever use of the sparsity of the parity-check matrix $H$.
For a binary LDPC code over binary-symmetrical channels, the error patten can often be identified by the bit-flipping algorithm.
Indeed, the positions to be flipped are exactly the error patten induced by the channel, i.e., $\mathcal{S}$, if the decoding is successful.
When one considers a $q$-ary LDPC code transmitted over a $q$-ary symmetrical channel, this error patten $\mathbf{e}$ can be efficiently
identified by verification-based message-passing decoder especially for large $q$\cite{SarvCS,ZhangqSc,BinGraphCS}. Indeed, both bit-flipping and verification decodings
can find their duals in reconstructing a sparse signal from sparse measurements\cite{XuExpanderCS,ZhangqSc}.

\subsection{Connection to Compressed Sensing}
Given the observation $\mathbf{s}\in \mathbb{R}^M$, the valid set of signal vectors
is, in fact, a coset of $C$ with the syndrome $\mathbf{s}$, namely,
\begin{equation}
  \label{eq:cs}
   \Lambda(\mathbf{s})=\left\{\mathbf{e}\in \mathbb{R}^N: H \cdot \mathbf{e} = \mathbf{s}\right\} = \mathbf{\hat{e}} + C,
\end{equation}
where $C=\{\mathbf{c}\in \mathbb{R}^{N}: H \cdot \mathbf{c} = \mathbf{0}\}$ and $\mathbf{\hat{e}}$ is one of coset leaders
with the minimum Hamming weight and the constraint of $H \mathbf{\hat{e}} = \mathbf{s}$.
In general, the number of coset leaders may not be unique.

The inherent connection between CS and linear codes over real numbers has been well exploited in \cite{ZhangCC_CS}.
With a coset representation of the valid signal set (\ref{eq:cs}), the connection can be further stated as follows.

\newtheorem{lem1}{Theorem}
\begin{lem1}
Let $H\in \mathbb{R}^{M\times N}$ be a measurement matrix and $\mathbf{e}$ is the sparse signal of length $N$. Further, assume that the $M$ measurements $\mathbf{s} = H \cdot \mathbf{e}$ and that $\mathbf{e}$ has at most $K$ nonzero elements, i.e., $\|\mathbf{e}\|_0 \leq K$. Then the syndrome decoder can properly recover the original signal $\mathbf{e}$
iff the number of coset leaders with the minimum Hamming weight equals 1. A sufficient condition for proper recovering is
that the minimum Hamming distance of the code $C=\{\mathbf{c}\in \mathbb{R}^{N}: H \cdot \mathbf{c} = \mathbf{0}\}$ satisfies $d_{\min}\ge 2K+1$.
\end{lem1}
\begin{proof}
The sufficient condition for proper recovering can be deduced as follows. If there is another solution $\hat{\mathbf{e}} \neq \mathbf{e}$,
we have that $\|\hat{\mathbf{e}}\|_0 \leq K$ and $\mathbf{s} = H \cdot \hat{\mathbf{e}}$.  Hence, $\hat{\mathbf{e}} - \mathbf{e}$ is a codeword of $C$ and
$\|\hat{\mathbf{e}} - \mathbf{e}\|_0 \ge d_{\min}$, which contradicts with $d_{\min}\ge 2K+1$.
\end{proof}

\section{Compressed Sensing with Incremental Sparse Measurements}
\subsection { Verification Decoding with Fixed Sparse Measurements}
For sparse measurements, LDPC matrices have been extensively employed for sparse measurements \cite{ZhangCSIT,BinGraphCS,LdpcCS}. It was shown in \cite{LdpcCS} that parity-check matrices of ¡°good¡± LDPC codes can be used as provably ¡°good¡± sparse measurement matrices under basis pursuit. In \cite{SarvCS,ZhangCSIT}, the verification decoder was shown to perform well with sparse measurements. Density evolution analysis of verification decoder in compressed sensing was reported recently in \cite{DEVerfCS}.

As a suboptimal syndrome decoder for linear codes over real numbers, the verification decoder with fixed sparse measurements (i.e., the sparse measurement matrix $H$ is fixed) can be described as follows:
\begin{enumerate}
\item  If a measurement is zero, then all neighboring variable nodes are verified as zero.

\item  If a check node is of degree one, then verify the variable node with the value of the measurement.

\item  If two check nodes overlap in a single variable node and have the same measurement
value, then verify that variable node to the value of the measurement.

\item  Remove all verified variable nodes and the edges attached to them by subtracting out the verified values from
the measurements.

\item  Repeat steps 1-4 until decoding succeeds or makes no further progress.
\end{enumerate}

Now, let us consider what happens whenever the verification decoder cannot make further progress (or simply converge).
When the verification decoder converges after several iterations, the variable nodes $\mathcal{V}$ can be partitioned into two disjoint sets, namely,
 the set of identified nodes $\mathcal{V}_I$ and the set of unidentified nodes $\mathcal{V}_U$ with
$\mathcal{V}= \mathcal{V}_I \cup \mathcal{V}_U$. If the decoder fails to recover the signal,  it is of high probability that $|\mathcal{V}_U|>0$.

\subsection { Verification Decoding with Incremental Sparse Measurements}

In practice, when the sparsity of the signal can be well estimated,  it is up to the number of sparse measurements for proper recovering. However, when the sparsity is under-estimated, the verification decoder may fails and more measurements are required. In general, one can employ incremental strategy for sparse measurements, which requires the design of rate-compatible sparse matrices just like one encounters in incremental redundancy hybrid automatic repeat request (ARQ) schemes. With verification decoding, one may employ the binary parity-check matrices developed for rate-compatible LDPC codes\cite{HaRCLDPC}.

In this paper, we, however, did not consider this approach but resort to a much simpler approach.  In simulations of verification decoding, it was noted that the number of unidentified variables after the verification decoder converges is often very limited compared to the signal length $N$, i.e., $|\mathcal{V}_U| \ll N$. Hence, it is now natural to employ an incremental measurement strategy with additional measurements tuned directly to the samples with determined locations in $\mathcal{V}_U$.

Here, we propose a two-step approach for the verification decoding with incremental measurements. In its first step, a sparse measurement matrix $H=[h_{m,n}]$ is employed with $M$ measurements. If the decoder fails, possibly due to an under-estimation of the sparsity, it initiates an incremental approach with direct sampling.

\begin{algorithm}
    \caption{Verification Decoding with Incremental Measurements}
\end{algorithm}
\begin{enumerate}

\item[-] Definition:
    \begin{description}
        \item[$k$]: iteration counter;
        \item[$\kappa_0$]: the threshold for the number of iterations;
        \item[$l$]: incremental measurement couter;
        \item[$\iota_M$]: the maximum number of measurements.
    \end{description}

\item[-] Initialize: \\ set $k=1$, $l=0$, $\mathcal{V}_I=\emptyset$.

\item  If $s_m=\sum_n h_{m,n} e_n = 0$ for some $m\in [1,M]$, its neighboring variable nodes are all identified as zeros, i.e,
$e_n=0, \forall n\in \mathcal{N}(m)$. Update the set of identified variable nodes as $\mathcal{V}_I\leftarrow \mathcal{V}_I \bigcup \mathcal{N}(m)$.

\item If $d_c(m)=1$ for some $m \in [1, M]$ and assume that $\mathcal{N}(m)=\{n\}$, its neighboring single variable node can be identified as $e_n = s_n$. Update $\mathcal{V}_I$ as $\mathcal{V}_I\leftarrow \mathcal{V}_I \bigcup \mathcal{N}(m)$.

\item For each $n\in [1,N]$, search over its neighboring check nodes to check if there exists two different check nodes $ m_1,m_2\in \mathcal{M}(n), m_1\neq m_2$ with equal but nonzero measurements $s_{m_1}=s_{m_2}\neq 0$. If yes, the variable node $n$ is identified as $e_n = s_{m_1}$ and other variable nodes are identified as  $e_{n'}=0, \forall n'\in \mathcal{N}(m_1) \bigcup \mathcal{N}(m_2)-\{n\}$. Update $\mathcal{V}_I$ as $\mathcal{V}_I\leftarrow \mathcal{V}_I \bigcup \mathcal{N}(m_1) \bigcup \mathcal{N}(m_2)$.

\item Remove all identified  variable nodes and the edges attached to them by subtracting out the identified values from
    the measurements.

\item If $k>\kappa_0$, go to Step 6). Otherwise, go to Step 7).

\item If $l<\iota_M$, locate the check node $m$ of minimum degree (choose any one of them if there are multiple such check nodes). Randomly choose an unidentified variable node $n$ neighboring to $m$, i.e., $n \in \mathcal{V}_U \cap \mathcal{N}(m)$ and directly sample the signal at this location. Update $\mathcal{V}_I$ as $\mathcal{V}_I \leftarrow \mathcal{V}_I \bigcup \{n\}$ and increase $l$ by 1.

\item Increase $k$ by 1, and repeat Steps 1)-6) until either all the variable nodes are identified
or the maximum number of iterations is reached.

\end{enumerate}

\section{Simulation Results}
In this section, we provide the simulation results for the proposed verification decoding with incremental measurements.
For source sparse signals, we consider the Gaussian sparse case where the entries of the signal are either 0 or a Gaussian random variable with zero
mean and unit variance.

We adopt the binary MacKay-Neal LDPC matrices \cite{MacKay-Neal} as sparse measurement matrices. Two LDPC code matrices are employed,
one of which is with the size of $M\times N = 738 \times 4095$, and the other is with the size of $M\times N = 2131 \times 16383$.
Here, $M$ can be seen as the number of initial sparse measurements. It should be noted that this matrix is constructed without 4-cycles.
Hence, the enhanced mechanism proposed in \cite{BinGraphCS} for further improving the performance of the verification decoding is not adopted.

\begin{figure}[htb] 
   \centering
   \includegraphics[width=0.5\textwidth]{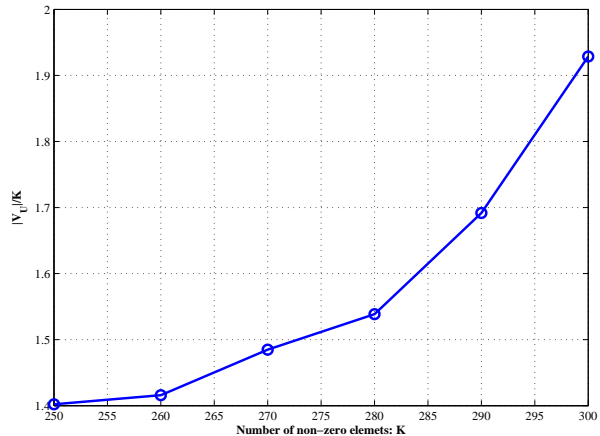} 
   \caption{ $|\mathcal{V}_U|/K$ versus $K$ conditioned on the decoding failure}
   \label{fig:1}
\end{figure}

\begin{figure}[htb]
   \centering
   \includegraphics[width=0.5\textwidth]{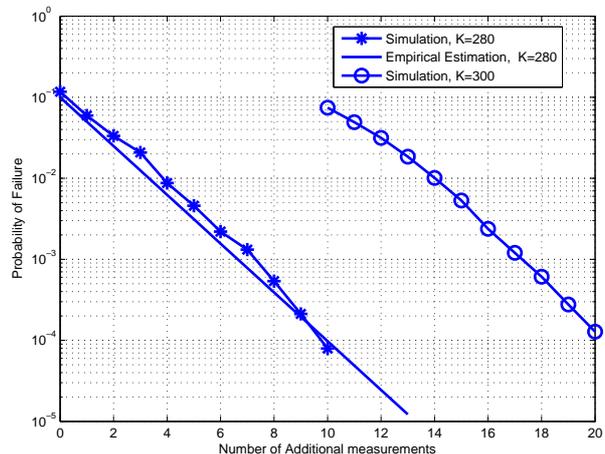}
   \caption{Probability of decoding failure against the number of additional measurements ($N=4095$, $M=738$).}
   \label{fig:2}
\end{figure}

\begin{figure}[htb]
   \centering
   \includegraphics[width=0.5\textwidth]{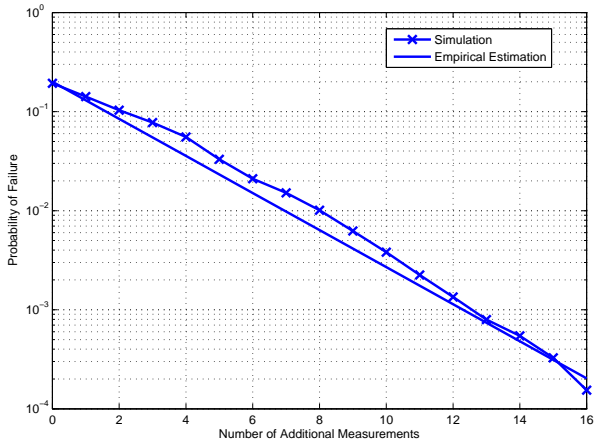}
   \caption{Probability of decoding failure against the number of additional measurements ($N=16383$, $M=2131$, $K=800$).}
   \label{fig:3}
\end{figure}

\subsection{Distribution of $|\mathcal{V}_U|$ Conditioned on the Decoding Failure}
The size of unidentified variable nodes is investigated whenever the verification decoder fails to recover the original sparse signal. Here, we focus on the measurement matrix of $M\times N=738\times 4095$.

The decoding failure occurs more frequently  when the number of non-zeros elements $K$ increases. Indeed, the probability of decoding failure is $P_f=1.47e-006$ when $K=250$ while it soon increases to $P_f=0.9$ when $K=300$. In Fig. \ref{fig:1}, we show the value of $|\mathcal{V}_U|/K$ versus $K$ conditioned on the decoding failure, which is averaged over 1000 decoding failures. As shown, much information about the sparse signal can be well retrieved even when the decoder fails to recover the original sparse signal.  In most cases, the size of unidentified variable nodes is greatly reduced compared to the length of the source signal $N$. Hence, it is of helpful for the decoder to further use this information when additional measurements are allowed to recover the signal.

\subsection{Decoding Performance Under Incremental Measurements}
If the number of sparse measurements keeps fixed, the probability of successful recovering the signal decreases as the number of non-zero elements $K$ increases.
In this subsection, we investigates the number of additional measurements for reducing the probability
of decoding failure when the sparsity of the signal keeps fixed.

As shown in Fig. \ref{fig:2}, the probability of failure decreases rapidly with the number of incremental measurements.
For $K=280$, the additional 10 measurements can reduce $P_f$ from about $10^{-1}$ to below $10^{-4}$.
For $K=300$, it clearly requires more measurements. Indeed, about 10 additional measurements are required for $P_f$ reaching about $10^{-1}$.
Once the probability of failure $P_f$ reaches around $10^{-1}$, the further employment of 10 measurements can again reduce $P_f$ to around $10^{-4}$.
Therefore, one could deduce that the additional one measurement can reduce $P_f$ to half of it. Let the number of additional measurements be $L$,
the probability of failure approximately obeys the following empirical estimation rule
\begin{equation}
  \label{eq:2}
   P_f(L) = P_f(0) \alpha^{-L},
\end{equation}
when $P_f(0)$ is around $10^{-1}$. The parameter $\alpha$ seems to be related to both the source sparse signal and the underlying sparse measurement matrix. As shown in Fig. \ref{fig:2},
 the empirical estimation with $\alpha=0.5$ is plotted. The empirical estimation is also plotted with $\alpha=0.65$ in Fig. \ref{fig:3}
 for the sparse matrix of size $M\times N = 2131 \times 16383$ and the sparsity of the signal is $K=800$. As shown, the empirical estimation coincides well with the simulation results.
 The exponential decay of the probability of decoding failure validates the efficiency of the proposed algorithm with incremental  measurements.

\section{Conclusion}

We have proposed a verification-based reconstruction algorithm with incremental sparse measurements.
With incremental measurements tuned to the sparse signal at the unidentified positions,
it has been proven very efficient for reducing the probability of decoding failure. With an additional direct-sampling approach, the implementation complexity is very low.

\section*{Acknowledgment}
This work was supported in part by the National Science Foundation of China under Grants 61032004, 61271335. The work of Yang was also supported by the National Science and Technology Major Project under grant 2010ZX0 3003-003-02, and by the National Basic Research Program of China (973 Program) under grant 2011CB302903.

\begin{thebibliography}{10}
\providecommand{\url}[1]{#1}
\csname url@samestyle\endcsname
\providecommand{\newblock}{\relax}
\providecommand{\bibinfo}[2]{#2}
\providecommand{\BIBentrySTDinterwordspacing}{\spaceskip=0pt\relax}
\providecommand{\BIBentryALTinterwordstretchfactor}{4}
\providecommand{\BIBentryALTinterwordspacing}{\spaceskip=\fontdimen2\font plus
\BIBentryALTinterwordstretchfactor\fontdimen3\font minus
  \fontdimen4\font\relax}
\providecommand{\BIBforeignlanguage}[2]{{%
\expandafter\ifx\csname l@#1\endcsname\relax
\typeout{** WARNING: IEEEtran.bst: No hyphenation pattern has been}%
\typeout{** loaded for the language `#1'. Using the pattern for}%
\typeout{** the default language instead.}%
\else
\language=\csname l@#1\endcsname
\fi
#2}}
\providecommand{\BIBdecl}{\relax}
\BIBdecl

\bibitem{DonohoCS}
D.~L. Donoho, ``Compressed sensing,'' \emph{{IEEE} Trans. Inf. Theory},
  vol.~52, pp. 1289--1306, Apr. 2006.

\bibitem{GilbertCS}
A.~Gilbert and P.~Indyk, ``Sparse recovery using sparse matrices,''
  \emph{Proceedings of the IEEE}, vol.~52, pp. 937--947, Jun. 2010.

\bibitem{XuExpanderCS}
W.~Xu and B.~Hassibi, ``Efficient compressive sensing with deterministic
  guarantees using expander graphs,'' in \emph{Proc. 2007 IEEE Inform. Theory
  Workshop., Lake Tahoe, CA,}, Sep. 2007, pp. 414--419.

\bibitem{ZhangCC_CS}
F.~Zhang and H.~D. Pfister, ``Compressed sensing and linear codes over real
  numbers,'' in \emph{Proc. 2008 Workshop on Inform. Theory and Appl., UCSD, La
  Jolla, CA,}, Feb. 2008, pp. 414--419.

\bibitem{ZhangCSIT}
------, ``Verification decoding of high-rate {LDPC} codes with applications in
  compressed sensing,'' \emph{{IEEE} Trans. Inf. Theory}, vol.~58, pp.
  5042--5058, Aug. 2012.

\bibitem{YihongCS}
Y.~Wu and S.~Verdu, ``Renyi information dimension: Fundamental limits of almost
  lossless analog compression,'' \emph{{IEEE} Trans. Inf. Theory}, vol.~56, pp.
  3721--3748, Aug. 2010.

\bibitem{SarvCS}
S.~Sarvotham, D.~Baron, and R.~G. Baraniuk, ``Sudocodes - fast measurement and
  reconstruction of sparse signals,'' in \emph{Proc. IEEE Int.Symp. Information
  Theory, Seattle, WA}, Jul. 2006, pp. 2804--2808.

\bibitem{BinGraphCS}
F.~Ramirez-Javega, M.~Lamarca, and J.~Villare, ``Binary graphs and message
  passing strategies for compressed sensing in the noiseless setting,'' in
  \emph{Proc. IEEE Int.Symp. Information Theory, Combridge, MA}, Jul. 2012, pp.
  1867--1871.

\bibitem{LdpcCS}
A.~G. Dimakis, R.~Smarandache, and P.~O. Vontobel, ``{LDPC} codes for
  compressed sensing,'' \emph{{IEEE} Trans. Inf. Theory}, vol.~58, pp.
  3093--3114, May 2012.

\bibitem{ZhangqSc}
F.~Zhang and H.~D. Pfister, ``Analysis of verification-based decoding on the
  q-ary symmetric channel for large q,'' \emph{{IEEE} Trans. Inf. Theory},
  vol.~57, pp. 6754--6770, Oct. 2011.

\bibitem{DEVerfCS}
Y.~Eftekhari, A.~Heidarzadeh, A.~H. Banihashemi, and I.~Lambadaris, ``Density
  evolution analysis of node-based verification-based algorithms in compressed
  sensing,'' \emph{{IEEE} Trans. Inf. Theory}, vol.~58, pp. 6616--6645, Oct.
  2012.

\bibitem{HaRCLDPC}
J.~Ha, J.~Kim, and S.~W. McLaughlin, ``Rate-compatible puncturing of
  low-density parity-check codes,'' \emph{{IEEE} Trans. Inf. Theory}, vol.~50,
  pp. 2824--2836, Nov. 2004.

\bibitem{MacKay-Neal}
D.~J.~C. MacKay and R.~M. Neal, ``Near shannon limit performance of low density
  parity check codes,'' \emph{Electronics Letters}, vol.~32, pp. 1645--1646,
  Aug. 1996.

\end{thebibliography}

\end{document}